\documentclass[12pt]{emulateapj}

\usepackage{epsf}
\usepackage{epsfig}

\begin{document}

\title{The signature of the Warm Hot Intergalactic Medium in WMAP and the
forthcoming PLANCK data.}

\author{
I. Suarez-Vel\'asquez\altaffilmark{1},
F.-S. Kitaura\altaffilmark{1}, 
F. Atrio-Barandela\altaffilmark{2},
J. P. M\"ucket\altaffilmark{1}}
\altaffiltext{1}{Leibniz-Institut f\"ur Astrophysik,
14482 Potsdam, Germany; isuarez,kitaura,jpmuecket@aip.de}
\altaffiltext{2}{F{\'\i}sica Te\'orica, Universidad de Salamanca,
37008 Salamanca, Spain; atrio@usal.es}

\begin{abstract}
We compute the cross-correlation between the Warm-Hot Intergalactic Medium 
and maps of cosmic microwave background temperature anisotropies using a 
log-normal probability density function to describe the weakly non-linear 
matter density field. We search for this contribution in the data measured by the 
{\it Wilkinson Microwave Anisotropy Probe}.  We use a template of projected matter 
density reconstructed from the Two-Micron All-Sky Redshift Survey as a tracer 
of the electron distribution.  The spatial distribution of filaments is modeled
using  the recently developed Augmented Lagrangian Perturbation Theory.
On the scales considered here, the reconstructed density field is very well 
described by the assumed log-normal distribution function.
We predict that the cross-correlation will have an amplitude of 
$0.03-0.3\mu$K. The measured value is close to $1.5\mu$K, compatible with 
random alignments between structure in the template and in the temperature
anisotropy data. Using the W1 Differencing Assembly to remove this systematic 
gives a residual correlation dominated by Galactic foregrounds. {\it Planck} could 
detect the Warm-Hot Medium if it is well traced by the density field reconstructed 
from galaxy surveys. The 217GHz channel will allow to eliminate spurious 
contributions and its large frequency coverage can show the sign change from 
the Rayleigh--Jeans to the Wien part of the spectrum, characteristic of the 
thermal Sunyaev-Zeldovich effect.
\end{abstract}


\section{Introduction.}

The baryon budget in the local universe shows a deficit relative to the 
predicted density synthesized in the big-bang (Fukugita et al. 1998; 
Fukugita \& Peebles, 2004). Galaxies and clusters contain about 10\% of the 
total number of baryons and an extra 5\% could be in the form of circumgalactic
medium (CGM) around galaxies, although the results of Gupta et al. (2012) of a 
large-scale massive hot gaseous halo around the Galaxy have been disputed
by Wang \& Yao (2012). 
Of the reminder 85-90\%, only half has been accounted for in the low redshift 
intergalactic medium (Danforth \& Shull 2008). Hydrodynamical simulations predict 
that the rest could reside within mildly-nonlinear structures with temperatures 
$0.01-1$KeV, called  Warm-Hot Intergalactic Medium (WHIM). The baryon fraction 
in this medium could be 40\% (Cen \& Ostriker 1999; Dav\'e et al. 1999, 2001, 
Smith et al. 2011). In the X-ray, the WHIM signature has been searched both 
in emission and in absorption. Soltan (2006) looked for the extended soft 
X-ray emission around field galaxies but his task was complicated by the need 
to subtract all systematic effects that could mimic the diffuse signal. 
The recent observational effort has concentrated in searching for absorption 
lines due to highly ionized heavy elements from the far-ultraviolet to the 
soft X-ray (see Shull et al. 2012, for a review). 
Alternatively, as the WHIM is highly ionized, in 
Atrio-Barandela \& M\"ucket (2006) and Atrio-Barandela et al. (2008) we
suggested that it would generate measurable temperature anisotropies
on the cosmic microwave background (CMB)
due to the thermal and kinematic Sunyaev-Zeldovich effect (TSZ and KSZ, 
respectively; Sunyaev \& Zeldovich 1970, Sunyaev \& Zeldovich 1972).
Our expectations were confirmed by Hallman et al. (2007) who found,
using numerical simulations, that after the contribution of resolved clusters
is removed, about one-third of the SZ flux from unresolved sources would be 
generated by unbound gas. More recently, Lieu \& Duan (2013) suggested 
that the line-of-sight column density of the ionized baryons in the local 
universe could be determined by monitoring quasar light curves.

Our model assumes that the undetected baryon phase, in the form of 
filaments of hot and low density IGM, is well described
by a log-normal probability distribution function. The filamentary
structure of the intercluster medium has been recently confirmed
using observations on interacting clusters. A joint analysis 
of ROSAT X-ray and Planck CMB data has provided the 
first detection of hot and diffuse intercluster gas, extending beyond the 
virial region of the cluster pair A399-A401 (Planck Collaboration, 2013).
If the WHIM can be traced with templates constructed from the matter
distribution, then cross-correlating those templates with CMB maps
could provide further evidence of the existence of the WHIM or
provide strong constraints on its spatial distribution. Based on
our log-normal model, in this article we compute the cross-correlation 
between CMB data with a template that traces the WHIM distribution. If 
galaxies from the Two-Micron All-Sky Redshift Survey (2MRS) are good tracers
of intergalactic gas, we can predict theoretically the amplitude of the TSZ 
component due to the WHIM. We compare our prediction with the cross-correlation 
derived from {\it Wilkinson Microwave Anisotropy Probe} (WMAP) data
(Jarosik et al. 2011). We show that the 
cross-correlation of galaxy templates and CMB data is dominated by random 
alignments. The differences between channels are due to Galactic foreground
residuals and instrumental noise. 
Planck could provide the first detection of the WHIM with our method by
using the 217GHz channel to remove systematics. Briefly, in Section~2 we 
describe our model; in Section~3 we describe the construction of 
the matter templates, show that they are well describe by our
log-normal formalism and compute the theoretically expected amplitude 
of the cross-correlation; in Section~4 we particularize our methods to
a matter reconstruction of the 2MRS catalog and 
WMAP data and discuss the prospects of a successful detection using the
forthcoming Planck data; finally, in Section~5 we present our main conclusions. 

\section{A log-normal model of the Warm Hot Intergalactic Medium.}

In our model, baryons in the WHIM are distributed like a log-normal random 
field. The log-normal distribution was introduced by Coles \& Jones (1991) 
to describe the non-linear distribution of matter in the universe. In this 
approximation, the number density of baryons $n_B({\bf x},z)$ is given by
(Choudhury et al. 2001, Atrio-Barandela \& M\"ucket 2006)
\begin{equation}
n_B({\bf x},z)=n_0(z){\rm e}^{\delta_B({\bf x},z)-\Delta_B^2(z)/2},
\label{logn}
\end{equation}
where ${\bf x}$ denotes the spatial position at redshift $z$ and $|{\bf x}(z)|$ 
is the proper distance, $\Delta_B^2(z) = <\delta_B^2(\bf x,z)>$, with 
$\delta_B({\bf x},z)$ is the baryon (linear) density contrast,
$n_0(z)=\rho_B(1+z)^3/\mu_B m_p$ with $\rho_B$, $m_p$ are the baryon density and
the proton mass, respectively; finally, $\mu_B=4/(8+5Y)$ is the mean 
molecular weight of the IGM and $Y=0.24$ is the helium weight fraction.
The log-normal distribution has been found
to describe very well the matter statistics at scales larger than $7h^{-1}$Mpc 
based on the improved Wiener density reconstruction from the Sloan Digital 
Sky Survey (see Kitaura et al. 2009). The linear baryon power spectrum is
related to the DM power spectrum by (Fang et al. 1993)
\begin{equation}
P_B^{(3)}(k,z)=\frac{P_{\rm DM}^{(3)}(k,z)}{[1+k^2L_{cut}^2]^2} ,
\label{pk}
\end{equation}
where the cut-off length is the scale below which baryon density perturbations
are smoothed due to physical effects like Jeans dissipation or shock heating 
(Klar \& M\"ucket 2010); for the WHIM, shock heating is assumed to be 
the dominating process. At any given redshift the comoving scale $L_{cut}$ is 
determined by the condition that the linear velocity perturbation 
${\bf v}(\bf x,z)$ is equal to or larger than the sound speed 
$c_s=(k_BT_{IGM}(z)/m_p)^{1/2}$ at each redshift. Here $m_p$ is the proton mass
and $T_{IGM}$ is the mean Intergalactic Medium (IGM) temperature. At redshifts
$z\le 3$, $T_{IGM}=10^{3.6}-10^4$K and its variation with
redshift is small (Tittley \& Meiksin, 2007). In addition,
$L_{cut}\approx L_0(1+z)^{1/2}$, with $L_0$ a constant and if
$T\simeq 10^4$K at present then $L_0\simeq 1.7h^{-1}$Mpc.
Hereafter, $L_0$ will be parameterized by $T_{IGM}$ 
(see Suarez-Vel\'asquez 2013 for details).

The number density of electrons 
$n_e$ in the IGM can be obtained by assuming equilibrium between recombination 
and photo-ionization and collisional ionization. At temperatures in the range 
$10^5-10^7$K and density contrasts $\delta \le 100$, the gas can be considered 
fully ionized. The two-point correlation function of the spatial variations of 
the electron pressure is given by
\begin{equation}
C(\theta)= \int_0^{z_f}\int_0^{z_f}\langle 
S_1(\hat{x},z)S_2(\hat{x}',z')\rangle dz dz' .
\label{cfull}
\end{equation}
where the integration is along lines of sight separated by an angle $\theta$. 
The TSZ WHIM temperature anisotropy is $\Delta T=y_cG(\nu)$, with 
$y_c=k\sigma_T/m_ec^2\int T_e n_e dl$ the Comptonization parameter along 
the line of sight, $n_e$, $T_e$ the electron density and electron temperature,
$m_ec^2$ the electron annihilation temperature, $k$ the Boltzmann constant, 
$\sigma_T$ the Thomson cross section, $G(\nu)=(x\coth(x/2)-4)$, with $x=h\nu/kT_0$ 
the reduced frequency, $T_0$ the CMB temperature, $dl$ the line element 
along the line of sight and $dz$ its corresponding redshift interval. Then if
$S_1=S_2=(k\sigma_T/m_ec^2) n_eT_e(dl/dz)$, from eq.~(\ref{cfull}) we can 
obtain the contribution of the WHIM to the power spectrum of CMB temperature 
anisotropies as
\begin{equation}
C_\ell=2\pi\int_{-1}^{+1}C(\theta)P_\ell(\cos\theta)d\cos\theta
\label{cl}
\end{equation}
where $P_\ell$ is the Legendre polynomial of multipole $\ell$. In Atrio-Barandela 
\& M\"ucket (2006) we modeled the gas as a polytrope. The amplitude of the 
resulting power spectrum was strongly dependent on the polytropic index; 
for some model parameters, it would be larger than $100(\mu K)^2$. Recently,
Suarez-Vel\'asquez et al. (2013) have refined the model for the shock-heated 
IGM by deriving the relation
$T_e$ versus $n_e$ from a fit to the phase diagrams obtained in various hydrodynamical 
simulations. The fit $\log_{10}T_e = 8-2[\log_{10}(4+x^b)]^{-1}$ with
$b=\alpha+x^{-1}$ and $x=n_e/\bar{n}_{B}$ the electron density in units of the 
mean baryon density, reproduces well the phase diagram in Kang et al. (2005).
Then, the model is parameterized by the IGM temperature and equation
of state with parameters in the range $T_{IGM}=[10^{3.6},10^4]$K and $\alpha=[1,4]$.
For this parameterization, temperature anisotropies grow with increasing electron 
temperature and decreasing cut-off length $L_0$.

\section{Cross-correlation of Matter Density Templates and CMB maps.}

The search of the WHIM contribution to the temperature anisotropies of
the CMB was pioneered by Hern\'andez-Monteagudo et al. (2004). We correlated 
the first year of WMAP 1yr with templates of projected matter density 
constructed from the Two Micron All Sky Survey (2MASS) galaxy 
catalog; all significant TSZ contributions 
were originated by clusters of galaxies and no evidence of the WHIM was found. 
A second approach was tried by G\'enova-Santos et al. (2009, 2013), who used a Monte 
Carlo Markov Chain to find the contribution in the CMB power spectrum but the 
evidence was not statistically significative. In this article we revisit the 
cross-correlation approach using templates of the density field reconstructed 
from 2MRS by Kitaura et al. (2012). The reconstruction technique is based on a 
Bayesian Networks Machine Learning algorithm (the \textsc{kigen}-code) which 
self-consistently samples the initial density fluctuations compatible with the 
observed galaxy distribution and a structure formation model given by second 
order Lagrangian perturbation theory (2LPT). We have used the Augmented Lagrangian 
Perturbation Theory (ALPT; Kitaura \& Hess 2012) to perform constrained simulations 
from the initial conditions found with \textsc{kigen}. It improves previous 
approximations at all scales by combining 2LPT with the spherical collapse
model and shows a higher correlation with the N-body solution than previous
methods. This approach enables us to find non-linear structures like 
filaments in the cosmic web to great accuracy. The number of solutions compatible 
with the observations of a galaxy sample is degenerate due to shell crossing 
and redshift distortions and the method provides an ensemble of 
reconstructions useful to estimate the uncertainties associated with the 
technique and intrinsic to the data (see Kitaura 2012 for details). 
Work to perform self-consistent reconstructions implementing ALPT within the 
\textsc{kigen}-code is in progress (Hess et al. 2013 in preparation).

\begin{figure}[t]
\centering
\epsfxsize=.55\textwidth \epsfbox{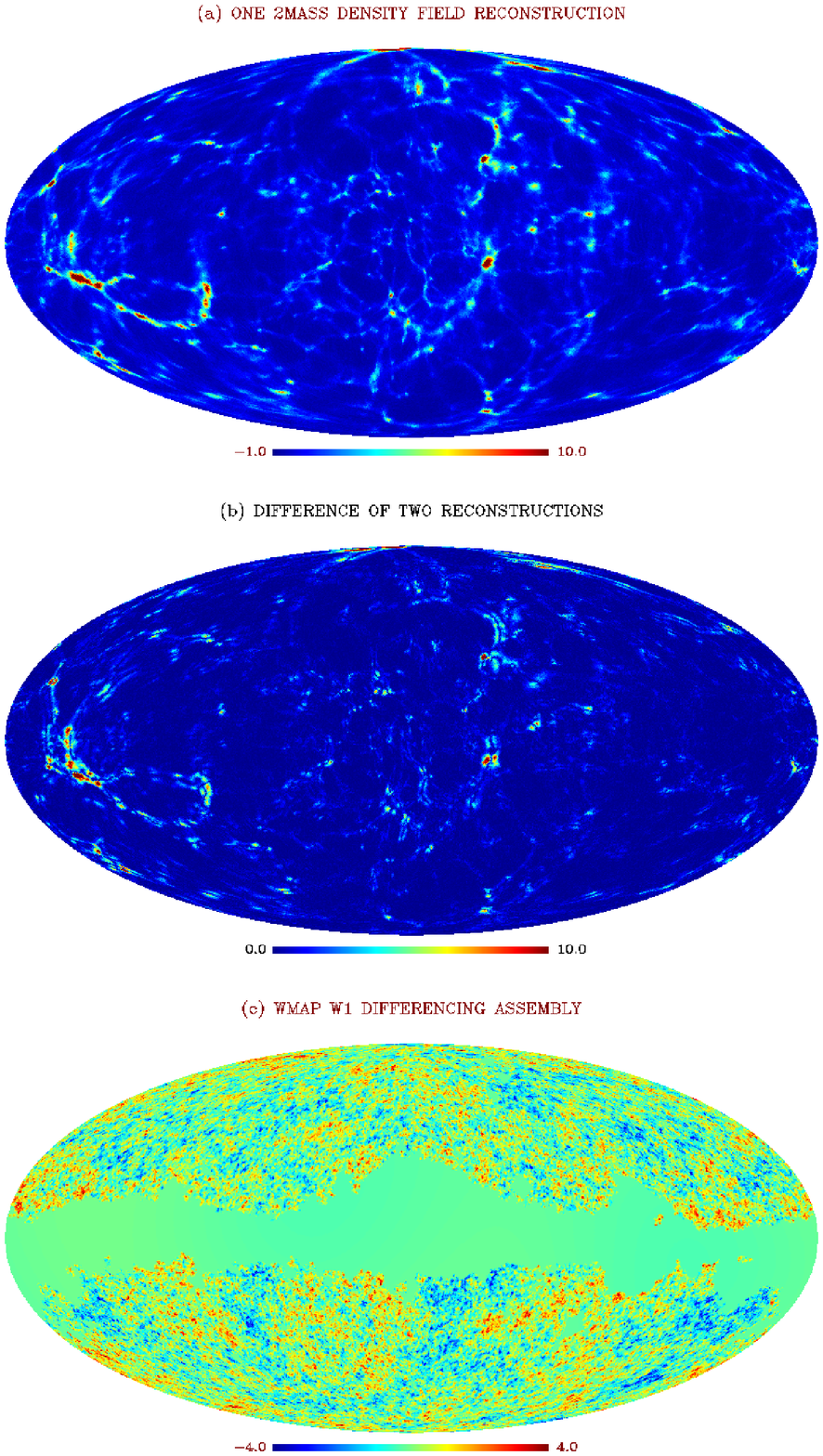}
\caption{\small Templates and CMB data at Healpix resolution
$N_{side}=128$. (a) Template $M$ of projected matter density
reconstructed from the 2MASS catalog, showing the filamentary structure
characteristic of the weakly non-linear regime.
(b) Difference of two reconstructed density fields showing small displacements
in the location of filaments and on the extent of massive structures. 
(c) WMAP data of the W1 Differencing Assembly normalized to 
zero mean and unit variance; the galactic plane has
been masked using the KQ75 galactic and point source mask.
}
\label{fig:maps}
\end{figure}

The reconstruction of the matter density field
requires evaluation of FFTs and, therefore, it is carried out on a cubic box.
Cubic boxes of side $160h^{-1}$Mpc and $180h^{-1}$Mpc  
were used to check  that the impact of boundary effects
in the density field is negligible on spheres of radius up to $80h^{-1}$Mpc.
To consider even  larger volumes one would require to model the 
``Kaiser Rocket effect'' in the selection function (Branchini et al. 2012) 
which is beyond the scope of this work. One such reconstruction,
denoted by $M$, is represented in Fig.~\ref{fig:maps}a. 
The matter distribution shows well defined filaments, characteristic for 
the mildly non-linear regime. The sky is represented using Healpix 
(Gorski et al. 2005) with resolution $N_{side}=128$, that corresponds
to an angular resolution of $27.5$\arcmin. We used a linear 
scale saturated at $500$ galaxies per pixel for better visualization.  
In Fig.~\ref{fig:maps}b we represent the difference between two reconstructions.  
While the cosmic web displayed in all reconstructions is very similar, the 
extension around very massive structures and the exact location of filaments 
differ from one reconstruction to another. For example, the uncertainty on 
the position of Coma is $2-3h^{-1}$Mpc, 
less than 5\% of its distance to the Local Group. Filaments that are slightly 
displaced appear in the difference map to be running side by side.
The ensemble of constrained simulations can be used to compute the mean and 
standard deviation of the density field in each cell, giving an estimate of 
the uncertainty in the position of the density peaks and filaments. 
For illustration, in Fig~\ref{fig:maps}c we represent 
the W1 Differencing Assembly of WMAP 7yr data. The galactic and point
source contaminations are masked using the extended temperature analysis WMAP mask KQ75.
The data is normalized to zero mean and unit variance outside the mask.

In Fig.~\ref{fig:power} we represent the mean power spectra of 24 different
template reconstructions of the matter distribution like in Fig.~\ref{fig:maps}a 
that are compatible with 2MRS (solid blue line) and the rms dispersion
around the mean (shaded area). The dot-dashed (green) line represents
the power spectrum of the projected density of 2MASS galaxies used in 
this reconstruction, corrected by inverse weighting with the selection function.
To avoid the instabilities due to small divisors, 37 pixels with number density
of galaxies equal to or larger than $10^3$ galaxies were eliminated.
All templates were normalized to zero mean and unit variance. 
Since the projected distribution of galaxies is close to a Poisson point process,
the power spectrum of 2MASS galaxies is roughly constant and very different,
at high multipoles, from the spectrum of the reconstructed matter density field.
At those scales, the matter distribution is smoother and the power falls.
The solid red line represents the power spectrum of the baryon 
density distribution computed by taking $S_1=S_2=n_e$ in eq.~(\ref{cfull}),  
with a cut-off length of $L_0\simeq 1.13h^{-1}$Mpc. We found that the best
fit to the power spectrum of the reconstructed template was obtained
when the integration in eq.~(\ref{cfull}) was restricted to the interval 
$60-120h^{-1}$Mpc, while the template of Fig.~\ref{fig:maps}a reproduces
the local volume out to $80h^{-1}$Mpc. This discrepancy is only apparent.
Even if the template has been reconstructed up to $80h^{-1}$Mpc, 
it does include higher modes since the method uses a $160 h^{-1}$Mpc box. 
Second, we have to take into account sampling variance; the local universe 
lacks power up to $30h^{-1}$Mpc (see e.g. Courtois et al. 2012). 
Spikes and oscillations in the spectra reflect the sample variance 
associated with observing a single universe. {\it While the distribution
of the WHIM and matter is different from that of galaxies at small scales,
they coincide at large scales, as it could be expected if galaxies traces
the overall matter distribution.} In the inset
we represent in a logarithmic scale the power spectrum at low multipoles 
to show that the reconstructed matter density field and the galaxy 
distribution on those scales are very similar. Lines follow the same
convention that in the main plot. To facilitate the comparison, the power 
spectrum of galaxies in this inset was multiplied by a factor of 20 to bring 
its amplitude closer to that of the power spectrum of the reconstructed matter 
density templates.

Except at very large scales, where the power spectrum is 
dominated by a few modes and sample variance is large, the templates of the 
reconstructed matter density field agree with the log-normal model at all angular scales.
This agreement demonstrates that templates like the one shown in Fig~\ref{fig:maps}a 
are very well described by our log-normal model, but it does not prove that the WHIM 
is stored in filaments. It only indicates that we can use the log-normal
model to predict the value of observable quantities that can be later
compared with the data. To this purpose, electron overdensities will be 
normalized to unit variance so the cross-correlation is given in units 
of temperature. For the template of Fig~\ref{fig:maps}a and for model
parameters in the fiducial ranges $T_{IGM}=[10^{3.6},10^4]$K 
and $\alpha=[1,4]$ the correlation at the origin is $0.03-0.3\mu$K. For 
a different parameterization that fits the Cen \& Ostriker (2006) phase 
diagram, the amplitude would be $0.01\mu$K, slightly lower. 

Contributions coming from gas in clusters or in the CGM are not included in 
our formalism. First, the log-normal model is restricted to overdensities 
$\delta\le 100$, while in clusters and in the CGM  are 
$\delta\sim 500-1000$ (Fukugita \& Peebles, 2004). Second, the distribution of 
galaxies or clusters is not well described by our log-normal model. For instance,
the CGM would induce anisotropies with the spatial distribution of galaxies 
and not filaments that, as Fig.~\ref{fig:power} shows, are very different.

\begin{figure}[t]
\centering
\epsfxsize=.55\textwidth \epsfbox{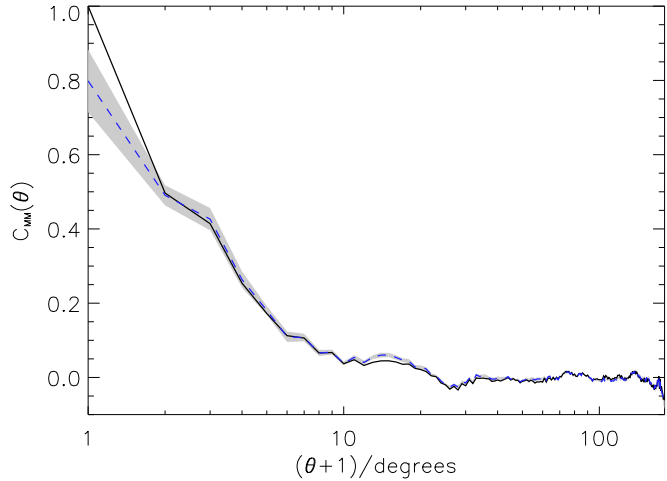}
\caption{\small 
Cross-correlation of the different reconstructions of the matter density field. 
The autocorrelation of the template M of Fig~\ref{fig:maps}a is represented by 
the solid (black) line while the blue dashed line and shaded are represents 
the mean and rms dispersion of the cross-correlation of M and the other 23
different reconstructions. For convenience, the values on the X-axis have 
been shifted by one degree.
}
\label{fig:corr_mm}
\end{figure}

If a template M is a good tracer of the gas
in filaments, cross-correlation with CMB data will give us an estimate
of the CMB temperature anisotropies generated by the WHIM measured,
for example, by WMAP. Since the reconstruction is not unique, we need 
to quantify the uncertainty introduced by using a template that does not 
exactly describe the distribution of baryons. To speed up the calculation
we degraded all our templates to a resolution of
$\sim 1.8^0$, corresponding to Healpix $N_{side}=32$.  At this resolution,
at separation of $1^0$ the correlation function is between first
neighbors, at $2^0$ between second neighbors, etc.
In Fig.~\ref{fig:corr_mm} we represent the cross-correlation of
the template M with the 24 different reconstructions of the matter density
field from 2MRS. The solid (black) line represents the autocorrelation function
of M; the dashed (blue) line represents the mean of the cross-correlation
of M with the other 23 different reconstructions and the shaded area is
the rms dispersion about the mean. At the origin, the autocorrelation
function and the mean of the cross-correlations differ by 20\%. Outside the 
origin, the autocorrelation is similar to the mean and very well within the 
$1\sigma$ error bar, indicating that the uncertainty on the position of the 
density peaks in the matter reconstruction of the 2MRS catalog is $\sim 2^0$. 
Fig.~\ref{fig:corr_mm} is very illustrative of the accuracy of our method.
Using resolution $N_{side}=32$ instead of $N_{side}=128$ we loose information 
about the correlation on scales $0.5^0-2^0$, but on those scales the 
reconstruction of the density field is uncertain
(see Fig~\ref{fig:maps}b). If we use a reconstruction of the density field that 
is not fully coincident with the true distribution of baryons in the local 
universe, we can expect to underestimate the true correlation function by 
the 20\% but the effect is negligible outside the origin. Then, by
restricting our analysis to Healpix resolution $N_{side}=32$ not only
the computation is faster, we also average over the uncertainties on 
our template that only contribute with a 20\% uncertainty on
the correlation at zero lag.
With this resolution there are no differences between WMAP 7yr and 9yr
data and the former were used.

\section{Application to WMAP Data}

\begin{figure}
\centering 
\plotone{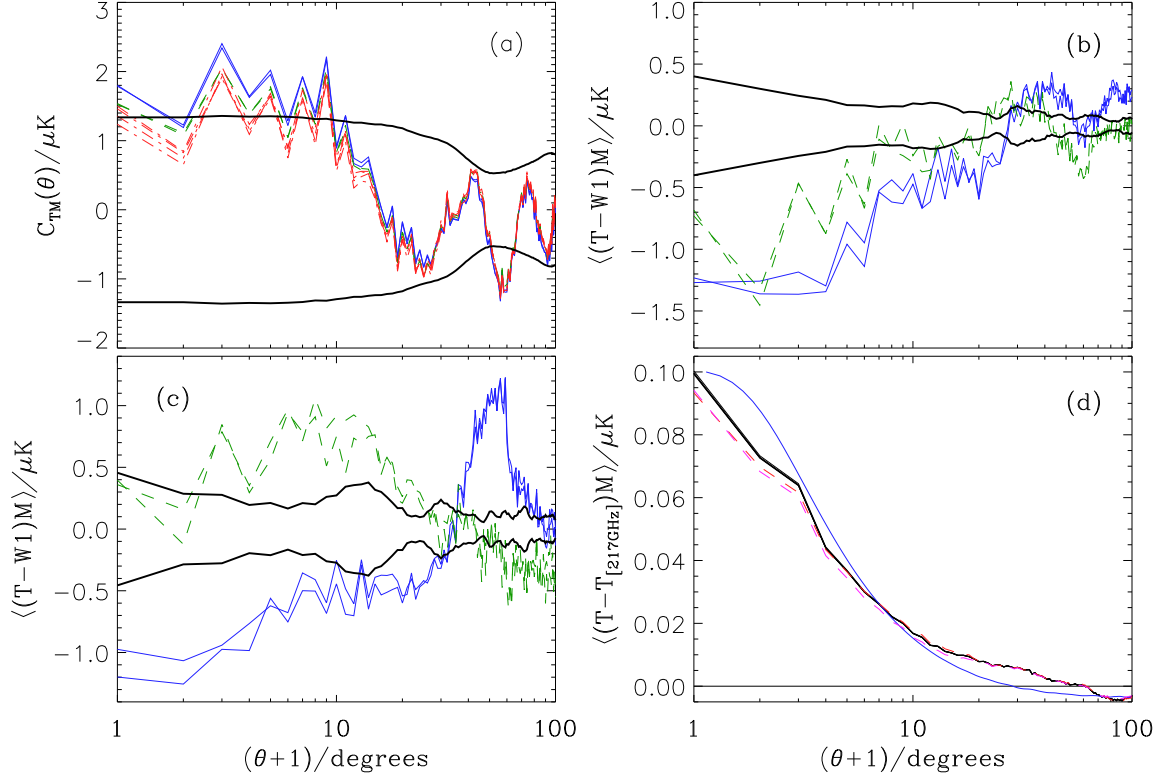}
\caption{\small (a) Cross correlation of the template $M$, represented
in Fig.~\ref{fig:maps}a, with the 8 DA of WMAP 7yr data. From top to bottom,
the solid (blue), dashed (green) and dot-dashed
(red) lines correspond to the Q, V and W channels.
The symmetric solid black lines correspond to the rms cross-correlation
of the template with 1000 randomly generated W1 maps.
(b) Cross-correlation of the Q (solid, blue) and V (dashed, green) channels after
subtracting the W1 cross-correlation, renormalized to account for
the removed TSZ contribution, i.e., $\langle (T-W1)M\rangle=[C_\nu(\theta)-C_{W1}(\theta)]/
[G(\nu)-G(W1)]$. Black lines represent the rms of the W2, W3, W4 DA
after subtracting the W1 cross-correlation. 
(c) As in (b) but the CMB-template correlation was computed 
only on pixels with $|b|\ge 30^0$. (d) Planck forecast: 
Solid (black) lines represent $\langle (T-T_{217GHz})M\rangle=
[C_\nu(\theta)-C_{217GHz}(\theta)]/[G(\nu)-G(217GHz)]$
for the 44-353GHz channels; the dashed (blue) line is the log-normal 
prediction. The dot-dashed lines (red and violet) represent
the cross-correlations of the template $M$ with Planck simulated data
containing a TSZ WHIM contribution with $T_e\propto const, n_e^2$ 
in the range $\delta\in[1,100]$, respectively.
Like in Fig~\ref{fig:corr_mm} the x-axis has been shifted by one degree.
}
\label{fig:data}
\end{figure}

Since the log-normal model describes reasonably well the power spectrum 
of the matter density then cross-correlation of a matter template M
with WMAP 7yr data would determine the fraction of gas traced by the matter.
In Fig.~\ref{fig:data}a we represent the cross-correlation of $M$ 
with the eight DAs of WMAP 7yr data. From top to 
bottom: solid (blue), dashed (green) and dot-dashed (red) lines correspond 
to the Q, V and W channels. The contribution of galactic foregrounds and point
sources were removed using the KQ75 mask  that eliminates 27\% of the sky
(see Fig~\ref{fig:maps}c). 
The matter density template $M$ was normalized to zero mean and unit variance
{\it outside the mask.} The symmetric solid (black) lines
represent the rms deviation of the cross-correlations of $M$ with
1,000 different random realizations of WMAP data that include cosmological
CMB signal and noise, but no foreground residuals. The
amplitude of the cross-correlation at the origin, $\sim 1.5\mu$K,
is much larger than the expected amplitude of $0.03-0.3\mu$K and it
is compatible with being due to random alignments between structures
in the CMB data and in the template. The differences
between the eight DA are due to the WHIM, noise and foreground 
residuals. Subtracting the correlation of M and W1 to the correlation of
M with the other seven DA removes the contribution due to random alignments
and yet, due to the frequency variation of the TSZ from 40 to 90 GHz, 
it leaves 14-18\% of the original TSZ signal in the V and Q bands, respectively. 
If in Fig~\ref{fig:data}a the differences were dominated by TSZ,
then the quantity $\langle(T-W1)M\rangle=[C_{\nu,M}-C_{W1,M}]/[G(\nu)-G(W1)]$ 
would give the correlation of the template with the map of the
Comptonization parameter, $y_c$-map. In Fig~\ref{fig:data}b we 
plot the $\langle M(y_c-map))\rangle$ correlation for WMAP.
The solid (blue) and dashed (green) lines corresponds 
to the Q and V channels, respectively. The differences between the
W2, W3, W4 DA with W1 would remove not only the intrinsic CMB, but also the 
foreground and WHIM signals. The rms variation of the subtracted correlations 
of the W channel provides a simple estimate of the error bar due to the instrumental 
noise, but does not account for foreground residuals and sampling variance 
contributions due to the variation of the number of pairs with
separation angle. The error bars are represented in Fig.~\ref{fig:data}b by 
the symmetrical solid (black) lines. The correlations 
$\langle(Q_{1,2}-W1)M\rangle$ and $\langle(V_{1,2}-W1)M\rangle$ are similar and
well outside the error bar, {\it but this is not evidence 
of WHIM.} In fact, the $\langle M(y_c-map)\rangle$ correlation
should be positive at the origin, not negative. We verify that this residual 
correlation is due to foregrounds by masking all CMB data with $|b|\le 30^0$. 
The results are presented in Fig.~\ref{fig:data}c.
Lines follow the same convention as in the panel (b). By restricting the
correlation to the polar caps, $\langle (Q_{1,2}-W1)M\rangle$ 
decreased by 30\% while $\langle (V_{1,2}-W1)M\rangle$ became positive, with
an amplitude of $0.4\mu$K at the origin, compatible with noise. Compared with 
Fig.~\ref{fig:data}b,
the correlation of the Q and V band are very different, reflecting the differences
in the foreground residuals between the $Q-W1$ and the $V-W1$ maps and an
increased sample variance as the fraction of the sky removed was larger.

The previous results indicate that WMAP is not well suited to
separate the TSZ contribution from that of the other components.
In this respect, Planck is a much better instrument. To 
forecast its performance, we simulated  6 Planck channels
with frequencies in the range $44-353$GHz. The simulated maps were
constrained to reproduce WMAP data for multipoles $\ell \le 256$.
The maps were downgraded to resolution $N_{side}=32$ so differences
in beamwidth and beam asymmetries have an unmeasurable effect.
The simulated data contain CMB and homogeneous white noise but did not
contain foregrounds, foreground residuals or (1/f) noise. 
At each frequency we added a WHIM component following the matter
distribution in $M$ assuming $T_e\propto n_e$
with a mean Comptonization parameter $\langle y_c-map\rangle=0.1\mu$K.
In the frequency range considered, the amplitude of
TSZ effect varies from $-1.9$ at 44GHz to $2.2$ at 353 GHz and passes
through the TSZ null at 217GHz. This channel was later used 
to subtract the correlation of $M$ with the CMB data due to random alignments.
In Fig.~\ref{fig:data}d we plot the results (solid black line), that 
are very close to each other and close to the theoretical expectation,
represented by the dashed (blue) line. The differences between the theoretical
curve and the simulated data are due to masking and pixelization. 
If foreground and noise inhomogeneities are not important and the template
traces the gas distribution, 
Planck will measure the WHIM TSZ correlation very accurately.

Together with the uncertainties in the reconstruction of the density field,
there are additional uncertainties associated with how well the galaxy 
template traces the electron pressure. The WHIM could be made of clumps,
with typical sizes $100h^{-1}$Kpc like the CGM. At $60h^{-1}$Mpc, a clump of this
size would subtend $6'$. The 217 and 343GHz Planck channels have the largest 
resolution of the instrument, $5'$; then, in Planck data we can neglect 
the effect of gas inhomogeneities at scales below $100h^{-1}$Kpc. Also, in our 
analysis we have assumed a smooth distribution of baryons and temperatures 
within the filaments.  As an average of the different Kang et al. (2005) models
we assumed the temperature scaling as $T_e\propto n_e$ so 
the electron pressure in the template scaled as $n_eT_e\propto n_e^2$. 
The simulations of individual filaments carried out in Klar \& M\"ucket (2010, 2012)
have characterized the state of the gas and its evolution inside the filament
in terms of the length of the initial perturbation. Their simulations 
indicated the existence of multiple phases and temperatures in the WHIM.  
To estimate the uncertainty associated with an inhomogeneous distribution
of gas and temperature within filaments, we repeat our simulations but added 
to the CMB data a TSZ contribution with two different temperature 
scalings in the range $\delta\in [1,100]$: (1) an isothermal gas $T\sim const$ 
and (2) a clumpier temperature distribution, $T_e\propto n_e^2$. 
In both cases, the mean amplitude of the $y_c$-map was $0.1\mu$K. 
These temperature--density relations can be considered extreme 
cases of the temperature variation in the IGM. Next, we cross correlate the
CMB data with our template $M$, where the electron pressure scales like 
$n_eT_e\propto n_e^2$. The resulting correlations are represented 
by dot-dashed (red and violet) lines in Fig~\ref{fig:data}d. The comparison of
these cross-correlations with our previous result shows that, at $N_{side}=32$
resolution, the shapes are similar and only differ by $\sim$5-10\% in the 
amplitude at the origin. In summary, the uncertainty on the 
cross-correlation of template and CMB data due to (1) how the template traces
the gas distribution and (2) how the electron pressure scales with density within
the template add at most an uncertainty of 30\% at the origin, being almost
negligible on scales above $2^0$.

The final error bar will certainly have non-zero contributions from 
foreground residuals, whose distribution and amplitude change with frequency,
and instrumental $(1/f)$ noise that will complicate the detection.
However, first and foremost, the TSZ has a distinctive signature,
different from that of any other foreground. It changes sign from the 
Rayleigh--Jeans to the Wien part of the spectrum. This change of sign 
will also be present in the correlation function, and could be sufficient 
to detect the effect. Second, foregrounds residuals are probably
largest close to the galactic plane, so computing the cross-correlation
at different galactic latitudes could help to disentangle the
different contributions, as for WMAP data.
Third, in our Planck simulations, the template contributes with
just $0.1\mu$K. According to our model, this is a reasonable expectation
for a large fraction of the parameter space. Naturally, the contribution
due a larger volume will be higher so templates constructed 
from deeper galaxy surveys would give statistically more significant 
detections. 

\section{Conclusions.}

We have computed the amplitude of the correlation between 
the WHIM with CMB data. We have shown that templates
of projected matter density that describe the weakly
non-linear regime are well represented by our log-normal model.
This permits us to predict that the amplitude of the cross-correlation
of the WHIM with CMB data is $0.03-0.3\mu$K depending on model parameters.
Assuming that the reconstructed matter density templates
trace the electron distribution, we 
computed the cross-correlation of the matter density field within a 
volume of $160h^{-1}$Mpc reconstructed using the 2MRS galaxy survey. 
The cross-correlation with WMAP data showed it to be dominated
by random alignments of CMB structures and galaxy filaments.
The differences between DA could be
due to the WHIM signal, but also to noise and foreground residuals.
We checked that foreground residuals are the most likely source of
the correlation by showing the results varied significantly when
removing data with $|b|\le 30^0$.

\begin{figure}[t]
\centering
\epsfxsize=.55\textwidth \epsfbox{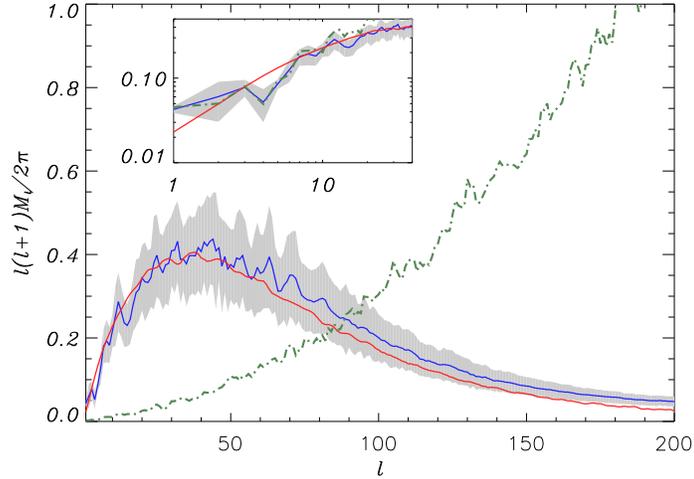}
\caption{\small Mean 
power spectra of the 24 reconstructed matter distribution templates including 
the one represented in Fig~\ref{fig:maps}a (solid blue line) and rms dispersion 
around the mean (shaded area). The dashed (green) line is the power spectrum of 
the density of 2MASS projected galaxies corrected by the selection function and 
the solid (red) line is the spectrum of baryon distribution given by our log-normal 
model.  All power spectra were normalized to zero mean and unit variance.
In the inset we represent the low order multipoles in a log-scale; 
the amplitude of the power spectrum of galaxies has been multiplied by a factor 20
to facilitate the comparison.}
\label{fig:power}
\end{figure}

Using 2MASS galaxies and WMAP data, Lavaux et al. (2013)
described evidence that baryons are distributed in galactic coronae.
This result is in apparent contradiction with Hern\'andez-Montegudo et al.
(2004) where cross-correlation of WMAP 1yr data with templates of 
projected galaxy density showed no evidence of ionized gas outside
known clusters of galaxies and, to some degree, with the results presented here.
Even if our matter density templates are different from a pure galaxy 
template their power spectra overlap (see Fig~\ref{fig:power}) and any 
signal traced by galaxies must also be traced at some level by our templates. 
In addition, the contribution due to random alignments, that we have 
shown to be dominant for WMAP, was not quantified. Due to the difference 
in methodology, our results are not fully comparable and consistency between 
both methods will increase the statistical significance of any detection.

Planck, with its large frequency coverage and high resolution 
is a more adequate instrument than WMAP to search for the WHIM
contribution. First, the 217GHz channel is very
close to the TSZ null, allowing to remove spurious correlations
that dominates the signal. Second, the instrument contains measurements 
on the Rayleigh--Jeans and Wien part of the spectrum. If the correlation is 
due to the WHIM, it will vary with frequency and change sign. 
Reconstructions of the matter density field using deeper surveys like 
the Sloan Digital Sky Survey will allow to probe the baryon distribution 
to higher depths, enhancing the signal. Alternatively, 
the reconstructed density field allows to select filaments aligned with
the line of sight, where the optical depth of the WHIM would be larger.
The higher resolution of Planck could facilitate to identify the WHIM
signal in those regions and clarify its spatial distribution, whether it is
distributed as a network of filaments or is stored in the galactic coronae.

\vspace*{1cm}
I.S.V. thanks the DAAD for the financial support, grant A/08/73458.
F.S.K. is a Karl-Schwarzschild fellow at the AIP.
F.A.B. acknowledges financial support from the Spanish
Ministerio de Educaci\'on y Ciencia (grants FIS2009-07238, FIS2012-30926
and CSD 2007-00050). He also thanks the hospitality of
the Leibniz-Institut f\"ur Astrophysik in the early
stages of this work.

\clearpage
\pagestyle{plain}

\end{document}